\documentclass{article}
\usepackage [urlbordercolor={1 1 1}] {hyperref}
\usepackage[none]{hyphenat}
\begin{document}
\author{Kirana Kumara P\\Centre for Product Design and Manufacturing,\\Indian Institute of Science,\\Bangalore, Karnataka 560012, India\\\textit{Email: kiranakumarap@gmail.com}
}
\title{A Study of Speed of the Boundary Element Method as applied to the Realtime Computational Simulation of Biological Organs}
\date{}
\maketitle
\begin{abstract}
In this work, possibility of simulating biological organs in realtime using the Boundary Element Method (BEM) is investigated. Biological organs are assumed to follow linear elastostatic material behavior, and constant boundary element is the element type used. First, a Graphics Processing Unit (GPU) is used to speed up the BEM computations to achieve the realtime performance. Next, instead of the GPU, a computer cluster is used. Results indicate that BEM is fast enough to provide for realtime graphics if biological organs are assumed to follow linear elastostatic material behavior. Although the present work does not conduct any simulation using nonlinear material models, results from using the linear elastostatic material model imply that it would be difficult to obtain realtime performance if highly nonlinear material models that properly characterize biological organs are used. Although the use of BEM for the simulation of biological organs is not new, the results presented in the present study are not found elsewhere in the literature.
\end{abstract}
\section{Introduction}
Realtime simulation of biological organs is a necessity while building realistic surgical simulators (realtime performance greatly enhances realism). Currently, realtime simulation in most cases is achieved using deformable models, spring-mass models, or models based on the Finite Element Method (FEM). There is lot of literature on the use of each of these approaches for the realtime simulation of biological tissues and organs including soft tissues (and organs that contain soft tissues, like liver). One can refer to any review paper or any of the many sources that contain lot of references to the relevant literature (e.g., [1-10]) to get an idea of the work that has already been done in this area.

Sources in literature widely agree on the fact that continuum mechanics based models (like models based on FEM) are desirable if one is interested to get accurate simulations but achieving simulations in realtime using the finite element models is difficult (if not impossible) because of the well known fact that realtime graphics needs about 30 computations per second whereas realtime haptics needs about 1000 computations per second.

One can see that the realtime simulation of three dimensional solids is a necessity for the continuum mechanics based realtime computational simulation of biological organs. One can also note that FEM is the most widely used numerical technique employed for this purpose. It is widely reported in literature that it is difficult to obtain realtime performance with FEM if nonlinear material behavior (that describes the material behavior of a biological organ, say liver) is to be incorporated. However it may be enough to describe biological organs with a linear elastostatic constitutive model sometimes, and in this case, it is reported in the literature that it is possible to perform simulations in realtime.

However there have been attempts to simulate realtime nonlinear behavior with FEM (e.g., [11, 12]). Information on the outcome of these attempts is not always readily available. Of course, FEM could be a great tool if one is not worried about the realtime performance; many commercial software packages (e.g., ANSYS) can easily perform a nonlinear simulation of biological organs if realtime performance is not a concern. Also, free and open source alternatives to the commercial finite element software packages are available, and one can use these packages to perform the simulations also; one can also use these packages as a background `engine' to perform the simulations `online' but it would be difficult to obtain the realtime performance still. Of course, there is no lack of references on the topic of simulation of biological organs where the realtime performance is not a concern; many a times, this is the case with surgical planning (e.g., [13]).

Apart from FEM, other numerical techniques like meshfree methods, and the Boundary Element Method (BEM) have been tried out in the literature for the realtime simulations. For example, author's work [14] uses the Finite Point Method (FPM) and precomputations to obtain realtime performance, whereas [15] uses the Method of Finite Spheres (MFS); both of these numerical techniques are meshfree methods. The review paper [1] identifies BEM as a candidate which could be better than FEM when it comes to the realistic and realtime simulation of biological organs for surgical simulations. References [16-18] use BEM for simulations. Reference [16] uses linear elastostatic boundary elements together with Woodbury formula; Woodbury formula is used to modify the characteristic matrix after the geometry is changed because of cutting. Reference [17] uses linear elastostatic boundary elements together with precomputations to achieve realtime simulations. Reference [17] also suggests parallel computing to obtain realtime performance but does not implement it. Reference [18] precomputes linear elastostatic boundary element solutions to obtain realtime performance. To simulate cutting, [18] uses interpolated pre-solutions which may not provide accurate solutions every time because cutting changes geometry and hence the solutions obtained from interpolated pre-solutions could be far from accurate at times. Reference [18] also suggests parallel computing to obtain realtime performance but does not implement it. A few more references that use BEM for simulations are mentioned in the next paragraph.

Reference [19] achieves realtime performance by using linear elastostatics and updating only part of the boundary element mesh where there is a change in the mesh, and updating the boundary element system of equations for this part of the geometry only. But this method may not be suitable every time especially when there is considerable difference between the original mesh and the modified mesh. Reference [20] deals with realtime boundary elements, but it is neither related to 3D linear elastostatics nor related to the simulation of biological organs. Reference [21] also uses linear elastostatics together with precomputations and interpolating precomputed solutions to achieve realtime performance. In this work also, upon cutting, boundary element mesh is updated only where there is a change in the mesh, and the boundary element system of equations is updated for this part of geometry only. A look up table is created in the precomputation step by applying unit displacement in three mutually perpendicular directions, on all the nodes in turn, and storing the solutions; since the principle of superposition holds good for linear elastostatic problems, the solutions stored in the look up table may be suitably superimposed to obtain the solutions during realtime simulations. In the concerned simulation, since the tumor is located completely within the soft tissue, [21] is an example where the boundary element has been used to solve a problem where a homogeneous and isotropic domain is enclosed within another homogeneous and isotropic region (i.e., a problem dealing with multiple regions or two regions); here, mesh that represents the tumor and the mesh that represents the soft tissue together form nested surfaces. Reference [21] also serves as an example where biological organs (soft tissue in particular) are modeled with the linear elastostatic constitutive law. Reference [22] deals with a hybrid deformable model and uses BEM together with a mass-spring model to simulate biological organs.

By carefully looking at the literature that deals with the application of BEM to the realtime simulation, one can observe that all of the works achieve the realtime performance by following one of these approaches or by following a combination of the following approaches: (i) Calculate the characteristic matrix for the system offline, and also take the inverse of the characteristic matrix offline (useful only in case of 3D linear elastostatics and also where the characteristic matrix does not change) (ii) Using a fine mesh only near the points where there is an interaction between the biological organs and the surgical tools (and also using a fine mesh where there is a contact, e.g., contact with the surrounding organs), and using a coarse mesh for the rest of the regions (iii) Whenever there is a change in the geometry, boundary element mesh is updated only where there is a change in the mesh, and the boundary element system of equations are updated for this part of geometry only (iv) Compiling a look up table during the precomputation step by applying unit displacement in three mutually perpendicular directions, on all the nodes in turn, and storing the solutions; since the principle of superposition holds good for linear elastostatic problems, the solutions stored in the look up table may be suitably superimposed to obtain the solutions during realtime simulations (v) Obtaining solutions by interpolating the solutions from the look up table (vi) Using hybrid models, e.g., using BEM together with a spring-mass model (vii) using Woodbury formula to modify the characteristic matrix after the geometry is changed (because of cutting, say).

One can see that directly and completely solving a BEM problem in realtime (without following any of the approaches (i) to (vii) above) definitely has advantages. For example, during surgical simulations, simulating cutting reduces to solving the BEM problem for the changed geometry. Also, simulating prodding, suturing, dealing with multiple regions, simulating the interaction between surrounding tissues etc. do not require any special techniques or any special approaches; they always reduce to just solving a linear elastostatic problem with different loads and boundary conditions (and for a different mesh sometimes). And since the time required to solve a linear elastostatic problem can be estimated beforehand (for a given number of total degrees of freedom, and a given element type), one can always be sure that the computations are performed within the allowed time limit (i.e., in realtime). One can also note that many of the sources in literature (including many of the ones mentioned above) recommend parallelization, but they mention it as `future work'. Author has not come across any source in the literature that reports the direct parallelization (i.e., without following any of the approaches (i) to (vii) above) of a BEM solution with the intention of obtaining realtime performance, although this approach has been recommended in many places.

Hence the present work aims to obtain the realtime performance using BEM. BEM code is parallelized and run on a computer cluster (up to 256 processors). This is a systematic study which demonstrates the applicability of BEM as applied to the realtime simulation of biological organs. Biological organs are assumed to be linear elastic, but in the end, the possibility of using BEM when nonlinear material behavior is desirable (e.g., in the simulation of soft tissues) is given a look. One can note that nobody has tried to perform nonlinear simulations in realtime using boundary elements so far, may be because on the one hand codes for nonlinear boundary element analysis are not readily available (forcing one to write ones own codes which would take considerable amount of time, and further, developing these codes is nothing but a huge developmental work which may need to be taken up by a team (not individuals)) and on the other hand `common sense' may be telling that it would not be possible to get realtime performance with nonlinear boundary elements because of inherent limitations and delays that may exist in computer systems. Also, it is important to note that when BEM is to be used to solve nonlinear problems, not only the boundary of the solution domain but also the volume of the solution domain needs to be discretized which means that one of the main advantages of using BEM over FEM is lost as soon as one tries to solve a nonlinear problem using BEM, and that may also be a reason why there has been no attempt to achieve the realtime performance while using nonlinear boundary elements.  

However, one can note that BEM is an established numerical technique that has already been used to solve a wide variety of problems including nonlinear problems. In the literature, one can find the application of BEM to problems which are more or less similar to the simulation of biological organs (e.g., [23]). But one can also note that these applications do not bother about realtime performance, and some of these bother about two dimensional (2D) applications only.

At this point, it is also informative to note a few points mentioned in the present paragraph. Studies on the applicability of FEM (not BEM) for the realtime simulation of biological organs are already available [24]. But such studies are not available for BEM. Since BEM has been identified in the literature as a good candidate for the simulation of biological organs, further studies on the usefulness of BEM for the simulation of biological organs is a necessity. Present work is a small step towards fulfilling that necessity. In contrast to the mainly analytical approach used in [24] to work out whether some simulation could be carried out in realtime, present work really carries out the simulations and then observes whether the simulations could be carried out in realtime. Although the approach followed in [24] has some advantages also, author can see that the approach followed in this work has the following advantages: (i) Even if it is found that some simulation cannot be carried out in realtime, simulation could be useful if one is happy with near realtime performance (although simulations are not strictly realtime)  (ii) Even if it is proved through some analysis that some simulations can be carried out in realtime, from the application point of view, one must really carry out the simulations and see that the simulations can be carried out in realtime (just an analysis is not sufficient but a demonstration is required).

One can observe that Reference [25] deals with the parallel implementation of the boundary element method for linear elastic problems on a MIMD parallel computer, but it deals with two dimensional (2D) problems only. Also this work does not deal with realtime simulations, and it does not worry about the applicability of BEM to the simulation of biological organs (realtime or otherwise).

Present work utilizes constant boundary elements of triangular shape. Next three paragraphs mention the advantages of using BEM, advantages of using (boundary) elements of triangular shape, and advantages of using constant (boundary) elements respectively. 

The BEM needs a meshing of only the boundary of any geometry (at least for linear problems). Hence lesser number of elements can describe geometry. Also, literature mentions that BEM shows very good scalability when parallelized to execute on a computer containing multiple processors; parallelizing the BEM would help in reducing solution times. BEM is generally thought to be an efficient numerical technique. In BEM, some of the unknowns can be displacements while at the same time the remaining unknowns can be tractions; hence computation of tractions from displacements is not needed; during surgical simulations, often the goal is to obtain reaction forces corresponding to prescribed values of displacements (i.e., goal is not to obtain displacements that correspond to known values of forces), and one can observe that BEM could be better than FEM for these problems. Often, one needs to know the solutions only on the boundary of an organ (or only at the nodes of boundary elements), and with BEM, there is no need to calculate solutions at internal points to get the solutions at the boundary, while FEM unnecessarily calculates solutions at internal nodes also. Also, it is widely mentioned in the literature that collision detection and rendering are easy with the type of object representation used in BEM. The system of equations resulting from BEM is dense while FEM produces sparse `characteristic' matrices. Since the algorithms and routines that take advantage of the sparseness of characteristic matrices are rare, BEM has an upper hand in this case (also since BEM needs meshing of only the boundary of the geometry and hence lesser number of elements can describe a geometry).

There are advantages in using boundary elements of triangular shape. Any complicated geometry can easily be represented by surface triangles. Also, one can construct a geometry using any standard CAD software package and then use the `stl export' option available in the package to obtain the surface mesh as an STL file (exporting a mesh as a VRML file may also serve the purpose). Mesh processing tools and remeshing tools (e.g., MeshLab, ReMESH) are widely available for surface meshes made up of triangles. Tools that can modify, edit, heal, and improve the quality of surface meshes that are made up of triangles (i.e., tools that can convert a mesh made up of ill shaped triangles to a mesh made up of well shaped triangles) are also widely available. One can note that it is easy to obtain 3D models described by surface triangles from 3D scan. One can also note that many of the software packages that can do 3D reconstruction of biological organs from 2D image sequences have the `stl export' option. Literature tells that it is easier to perform rendering and collision detection with meshes made up of surface triangles.

Now, advantages of using constant boundary elements are explained. Constant boundary elements are fast in the sense that they do not need complicated shape functions that are computationally expensive. Constant elements are easy to use and program. No connectivity information is needed if constant elements are used. This makes it easy to parallelize the code, and the resulting code is highly scalable. Constant elements are suitable for handling multiple regions since nodes are located completely within the elements (not on the edges or corners). Of course, accuracy may be poor for a given number of elements when compared to linear and quadratic elements. But for applications like the realtime simulation of biological organs, accuracy is not too important when compared to speed.

Present paper is organized as follows. Next section gives a description of hardware and software utilized while carrying out the present work. The subsequent section forms the core of the paper and it presents studies on the speed of BEM, as applied to realtime simulations. The last section concludes the paper while making some interesting observations.
\section{Descriptions of Hardware and Software Utilized}
Present work needs a BEM code whenever a solution is to be obtained through BEM. Now, the next few paragraphs give a brief description of the code used.

Whenever the BEM code is needed, present work makes use of a BEM code written from scratch by the present author; the code is not in whole or in part a copy or a translation of any of the codes developed by someone else; the code can be used to solve any three dimensional linear elastostatic problem (without considering body forces) using constant boundary elements. The code (which is free and open source software) is available for download from [26]. Present work does not attempt to explain the theory behind the code. One can refer to any standard text book (e.g., [27-30]) on boundary elements to know the theory behind boundary elements. Good tutorials like [31] may also serve as quick but effective introduction. However, the next few paragraphs give a very broad overview of the code.

The code addresses strong singularity by using `rigid body modes' and weak singularity is taken care of by utilizing higher number of integration points over each of the boundary elements. Now the next paragraph tells a few words about `rigid body modes'.

When an integral as encountered in the 3D linear elastostatics without body forces (with constant elements) is strongly singular, the value of the definite integral exists only in the sense of Cauchy Principal Value (CPV).  Cauchy Principal Values in this case may be found either by direct evaluation [32, 33] or by using `rigid body modes' explained in [28, 29]. The code uses `rigid body modes' or `rigid body considerations' as explained in [28, 29] to evaluate strongly singular integrals.

The code uses 16 by 16 numerical integration (i.e., 256 integration points over each element) by default. Higher number of integration points ensures accurate evaluation of weakly singular integrals. Also, the code has a provision to use 4 by 4, 8 by 8, or 32 by 32 integration points also; location of Gauss points and the corresponding weights for these cases are listed in the code itself; here the location of Gauss points and weights are obtained from [34]. Test runs showed that using 4 by 4 integration does not give very accurate results. Test runs using 8 by 8 integration gave accurate results but the code uses 16 by 16 integration by default just to ensure that one does not get erroneous results just because of inaccurate integration. It was also observed from test runs that there is not much improvement in accuracy by using more than 16 by 16 integration points. Of course, using 8 by 8 integration instead of the default 16 by 16 integration would definitely make the code execute faster.

The source [26] provides the code in three versions: (i) A MATLAB code for solving three dimensional linear elastostatic problems using constant boundary elements while ignoring body forces (ii) A Fortran translation of the MATLAB code (iii) A parallelized version of the Fortran code. In the present work, whenever a simulation needs to be run on a desktop computer (with or without using a GPU), the MATLAB version of the code is used. And whenever a simulation needs to be run on a computer cluster, the Fortran versions of the code are used. 

Author had made use of many resources while writing the code. A few of the many noteworthy ones are [35-41]; these resources could be of help if one is interested to understand the code fully. It is also noteworthy to mention that the author had made extensive use of the information contained in many of the online forums also, especially while running/compiling/developing the code.

In the present work, simulations are tried out on two hardware platforms. The first hardware platform is a desktop computer, and the second hardware platform is a computer cluster.

Coming to hardware and software that are used when simulations are carried out on a desktop, the MATLAB codes (with or without GPU) are run on a desktop computer (Intel(R) Xeon(R) CPU E5405 @ 2.00GHz (8 cores), 8GB RAM, Mainboard: Intel D5400XS, Chipset: Intel 5400B, Windows xp Professional x64 Edition, SSD: Corsair CSSD-F60GB2, MATLAB2011b (32 bit version), GPU: NVIDIA Quadro 4000 (Driver Version 6.14.12.9573)). Apart from the solid-state drive (SSD), the simulations were tried out using the ordinary hard disk also. Also, 64 bit version of MATLAB2011b was tried out. But since it was found from the simulation results that the SSD is about 1.5 times faster when compared to the conventional hard disk and the 32 bit MATLAB is about 10 times faster when compared to the 64 bit MATLAB, it is decided to use the SSD and the 32 bit MATLAB.

Coming to hardware that is used when simulations are carried out on a cluster, a computer cluster consisting of 17 nodes is used. The cluster consists of 9 nodes with 32 cores each (2.4 GHz AMD Opteron 6136 processor, 64 GB RAM) and 8 nodes with 64 cores each (2.2 GHz AMD Opteron 6274 processor, 128 GB RAM). A 500 GB SATA HDD (3 Gbps) is used for OS and system software. An Infiniband Card (MHQH19B-XTR) is used for MPI communication, and Dual-port Gigabit Ethernet Connectivity is used for enabling logins and NFS. And coming to software, Intel Composer XE (Version: 2011.5.220) which includes Intel Fortran compiler together with Intel Math Kernel Library (Intel MKL) is used with MVAPICH2 (Version: 1.8-r5423). CentOS 6.2 (Linux x86\_64 Platform) is the operating system, and the batch scheduler software `Torque' is used for job scheduling and load balancing. Whenever Torque is required, `mpiexec' (Release 0.84) from Ohio Supercomputer Center (OSC) is used instead of the `mpirun', in the job script. Although the cluster has 800 processors in total, only 256 cores are used in the present work. This is because the author's organization does not allow any individual to use more than 256 processors at any given point of time in the concerned computer cluster; this is to avoid a single user utilizing all the available computing resources which may cause problems to other users of the computer cluster. Also, from the results to be presented later in this paper, one can see that there is no need to go for more number of processors since it may not lead to any speed up.
\section{Studies on the Speed of the Boundary Element Method}
Many of the subsections in this section utilize a sample problem. The sample problem is about taking up a 4 mm by 4 mm by 4 mm cube, and completely fixing one face of the cube, and applying traction of 4 N/mm\textsuperscript{2} in the y-direction over the whole of the opposite face. Each face of the cube is discretized into sixteen boundary elements. Young's modulus is assumed to be equal to 200000 N/mm\textsuperscript{2}, and Poisson's ratio is assumed to be equal to 0.33; aim of the simulation is to obtain the displacements for the elements that are subjected to known tractions and also to obtain the tractions for the elements that are subjected to known displacements, using constant boundary elements, and using linear elastostatic assumption and ignoring body forces. To demonstrate the speed that can be achieved using BEM, one needs to solve a problem using BEM, and the sample problem is of help here. Whenever the phrase `sample problem' appears in the present paper, the phrase refers to this problem only.

The following subsections give information about speeds that can be achieved using different hardware and software; results indicate whether or not a particular hardware-software combination can offer realtime performance for simulations considered in a particular subsection.   

\subsection{Solving the Sample Problem on the Desktop Computer (Without Manual Parallelization)}
In this subsection, no GPU is used. Also, no manual parallelization is attempted. But of course, the automatic parallelization available in MATLAB is utilized.

Time needed to solve the sample problem is found using the MATLAB commands `tic' and `toc'. It was found that it took 2.103 s, 2.146 s, and 2.092 s respectively, during three trials, to solve the sample problem using the default option of using all 8 cores (i.e., using the default fully automatic parallelization available in MATLAB) in the desktop. Next, the problem is solved using a single core in the desktop, and the time needed to solve the problem in this case was found to be 2.101 s, 2.090 s, and 2.101 s respectively, during three successive trials. Now, one can see that it takes more time to solve the problem on 8 cores, when compared to the time needed to solve the problem on a single core. Hence, the automatic parallelization offered by MATLAB is not of use for this problem (in fact, in this case, performance offered by the automatically parallelized code is worse when compared to the performance offered by the sequential code), and one can also observe that the code takes about the same time to run on the 8 cores as it takes to run on a single core. Also, one can see that the simulation is far from being realtime.

Now a small note on the speed of MATLAB in general. There is a general opinion that MATLAB is slower when compared to `lower level' programming languages like Fortran, C, C++; in fact, there have been many discussions in online forums on topics like ``Speed of MATLAB versus speed of Fortran'' etc. Upon going through these forums, one comes across varying opinions like ``Present day compilers for high level languages such as FORTRAN are so good that the codes written in high level languages are almost as fast as the same codes written in an assembly language, at least when the programmer is not extremely skilled'', ``MATLAB used to be about 100 times slower when compared to languages like FORTRAN; but once MATLAB started using the modern JIT compiler, MATLAB is about 10 times slower'', ``MATLAB has many built-in functions for scientific applications and the functions are so optimized for speed that it is difficult to write the same functions oneself, in languages like FORTRAN, to achieve the same speed'', ``If the programmer is not skilled, languages like FORTRAN could be slower when compared to MATLAB'', or ``Whether Fortran is faster when compared to MATLAB is highly dependent on the problem in hand as well as the skill of the programmer''. In the present work, instead of assuming that the MATLAB is faster when compared to Fortran or otherwise, codes are written both in MATLAB and Fortran and whether Fortran is faster is decided based on the results from the actual runs; in fact, from the results presented (or to be presented) in this section, one can conclude that Fortran is significantly faster when compared to MATLAB; further, a Fortran code can be parallelized and run on a computer cluster whereas it is not easy to find a MATLAB version that can run on a cluster, and even if a MATLAB version that can run on a cluster is found, that version has its own limitations. Of course, people make their MATLAB codes faster by making use of `System Objects', `MATLAB Coder', `MEX functions' etc. (e.g., [42]), but the present work aims to achieve the realtime performance without making use of these specialized approaches; also, all simulations cannot be translated to these approaches, and one can note that the GPU implementations of these specialized features are not available often, and further, it may not be possible to achieve the realtime performance with MATLAB even after employing these specialized techniques.             

\subsection{Solving the Sample Problem on the Desktop Computer (with Manual Parallelization)}

Since it is found from the previous subsection that the realtime performance cannot be obtained through either a sequential or an automatically parallelized MATLAB code that solves the sample problem, attempt is made in this subsection to manually parallelize the code.

Now, only a portion of the MATLAB code is parallelized; idea is that if realtime performance can be obtained for this portion of the code, manual parallelization can be attempted for a larger portion of the code; and of course, there is no need to attempt to parallelize the whole MATLAB code if one cannot obtain the realtime performance even for a portion of the whole code. Manual parallelization is attempted only for the `for' loop that calculates the components of the unit normal to the element faces; this is because the concerned `for' loop is ``embarrassingly parallel'', and also because manual parallelization is as easy as just replacing `for' with `parfor'. One can note here that ``embarrassingly parallel'' is a terminology used by the High Performance Computing (HPC) community to indicate that the program to be parallelized is highly scalable. Ideally, the time required to solve an ``embarrassingly parallel'' problem reduces linearly as the number of processors increase (ignoring the time needed for inter-processor communications, and the linear scalability is possible only up to certain number of processors).  

The time taken to execute the loop after just replacing `for' with `parfor' (without initializing `matlabpool') is 0.102 s, 0.102 s, and 0.102 s, for the three trials considered; here, the code runs on the `client' only, not on the `MATLAB workers'.

Next, `matlabpool' is used to initialize `matlabpool', and `matlabpool close' is used to close `matlabpool'. Again, `for' is replaced with `parfor'. The time taken to execute the concerned `for' loop is 13.680 s, 13.667 s, and 11.533 s, for the three trials considered; times include the time taken to initialize and close `matlabpool'. If one ignores the time taken to initialize and close `matlabpool', the time taken to execute the concerned `for' loop is 0.324 s, 0.322 s, and 0.330 s, for the three trials. Hence one can see that initializing and closing `matlabpool' takes a lot of time. All the simulations mentioned in this paragraph use the default 8 `workers' (since there are 8 cores in the desktop); `workers' are also known as `labs'.

Now, simulations exactly similar to the ones in the last paragraph but only with 1 `worker' are carried out. The time taken to execute the same `for' loop, excluding the time taken for executing `matlabpool' and `matlabpool close', is 0.263 s, 0.264 s, and 0.265 s, for three trials.

Now, the time taken for the same simulation as the one in the last paragraph but with 2 `workers' is found to be 0.277 s, 0.274 s, and 0.275 s, for three trials. The time taken for the same simulation as the one in the last paragraph but with 4 `workers' is found to be 0.290 s, 0.290 s, and 0.289 s, for three trials.

One can observe that as the number of `labs' increase, simulation becomes slower in this case. Also, one can observe that none of the simulations mentioned in the present subsection could be completed in realtime. One can also observe that if one does not substitute `parfor' for `for', and uses the default automatic parallelization of MATLAB, the time taken to execute the `for' loop is just 0.032 s, 0.032 s, and 0.032 s, for the three trials, which makes the simulation a realtime one; the default option of using all the available 8 cores is used here. Instead of using the default option of using all the available cores, if `maxNumCompThread' is used to restrict the number of cores to be used, following is the time needed to execute the portion of the MATLAB code within and including the concerned `for' loop, without substituting `parfor' for `for': 0.032 s, 0.032 s, and 0.032 s for the three trials if only one core is used; 0.032 s, 0.032 s, and 0.032 s for the three trials if two cores are used; 0.033 s, 0.032 s, and 0.032 s for the three trials if four cores are used; 0.032 s, 0.032 s, and 0.032 s for the three trials if eight cores are used.   

From the results presented in this subsection, at least for the problem considered in this subsection, there is no use in manually parallelizing the MATLAB code using `parfor'. One should also note that not all statements that can be put inside a `for' loop can be put inside a `parfor' loop. Of course, one cannot rule out the possibility of a future version of MATLAB providing a better implementation of `parfor'.
    
\subsection{Solving the Sample Problem on a GPU}

Since the previous two subsections are not successful in obtaining realtime performance, there is a need to run the sample problem on the GPU to see whether one can achieve realtime performance.

GPU computing features available in MATLAB depend on the features available in CUDA and GPU drivers, which in turn depend on the features supported by the GPU hardware. But one has to note that only a small subset of MATLAB functions can be run on GPUs using the MATLAB Parallel Computing Toolbox. One can observe that newer versions of MATLAB (and the Parallel Computing Toolbox) have better support for GPUs, and one can definitely hope to see the future versions of MATLAB enabling more and more MATLAB functions to be run on GPUs using Parallel Computing Toolbox. But, as of now, a lot more needs to be done from the software developers to make many MATLAB functions to readily run on GPUs. Not only functions, but some MATLAB scripts cannot readily be ported to GPUs. Of course, programs written in lower level languages like Fortran and C may also be modified to run on GPUs; however, this task may not be easy always, and sometimes, porting a code to a GPU could itself be a research problem. Also, whenever a GPU is used, one needs to transfer the variables from CPU to GPU, and after performing computations on the GPU, results have to be transferred from GPU to CPU; and these data transfers are time consuming. Although author is aware of these limitations of GPU computing, present subsection makes an attempt to run the MATLAB code on a GPU with the intention of obtaining realtime performance, using the MATLAB Parallel Computing Toolbox.

Just like in the last subsection (which tried to parallelize the MATLAB code using `parfor'), only a portion of the MATLAB code that solves the sample problem is parallelized in this subsection first; idea is that if realtime performance cannot be obtained for even a portion (or a part) of the code, there is no need to attempt to parallelize the whole MATLAB code. Hence the parallelization is now attempted only for the `for' loop that calculates the components of the unit normal to the element faces. Also, whenever the parallelized code is run on the GPU, there is a need to transfer the variables from CPU to GPU.

With prior initialization of the GPU arrays using `parallel.gpu.GPUArray.zeros', time needed to execute the MATLAB program from the beginning of the program to the end of the concerned `for' loop is found to be 1.249 s, 1.247 s, and 1.251 s, for the three trials. But if GPU arrays are not initialized using `parallel.gpu.GPUArray.zeros', time needed to execute the MATLAB program from the beginning of the program to the end of the concerned `for' loop is found to be 1.789 s, 1.807 s, and 1.805 s for three trials.

The simulation that is exactly same as the one carried out in the last paragraph but carried out on the CPU alone (without using the GPU at all) takes 0.172 s, 0.173 s, and 0.173 s (for three trials) to complete, if run on a single core of the desktop; but if all the 8 cores of the desktop are utilized, the same simulation takes 0.173 s, 0.173 s, and 0.173 s, for three trials.  

Now, with prior initialization of the GPU arrays using `parallel.gpu.GPUArray.zeros', time needed to execute on the GPU the concerned `for' loop alone is 1.252 s, 1.251 s, and 1.253 s for three trials. Now, if the time needed to execute `parallel.gpu.GPUArray.zeros' is also taken into account, time needed to execute on the GPU the concerned `for' loop alone plus the time needed to execute `parallel.gpu.GPUArray.zeros' is 1.271 s, 1.258 s, and 1.266 s for three trials. But if GPU arrays are not initialized using `parallel.gpu.GPUArray.zeros', time needed to execute on the GPU the concerned `for' loop alone is 1.728 s, 1.708 s, and 1.729 s for three trials.

One can see that none of the simulations that are tried out in the present subsection turned out to be realtime. Also, as far as the problems considered in this subsection are concerned, there is no advantage in using a GPU instead of a CPU. And since it is found that not even a portion of the sample problem could be executed in realtime on a GPU, there is no point in trying to run the whole of the sample problem on a GPU in realtime.

\subsection{Using the GPU to Solve the System of Equations only}

Results presented in the previous subsections of the present section show that it is not possible to solve the sample problem in realtime by making use of a desktop computer loaded with MATLAB, even if a GPU together with the MATLAB Parallel Computing Toolbox is made use of. One can note that the sample problem is a small-sized problem and if it is not possible to solve this small-sized problem in realtime, it would not be possible to solve a larger sized problem in realtime. But it is of use sometimes, even if one manages to solve only a part of the sample problem in realtime. For example, if one is happy with linear elastostatics and if there is no change in the geometry during a simulation, the `characteristic matrix' and its inverse can be precomputed, and the problem then reduces to just a matrix multiplication; and in this case, if one can manage to complete the matrix multiplication in realtime, it could be as good as solving the whole problem in realtime. Hence there is a need to see whether particular portions of the sample problem can be executed in realtime, either by making use of a GPU or not.

Hence, in this subsection, attempt is made to obtain the realtime performance while solving a system of linear simultaneous algebraic equations; also, in the next subsection, attempt is made to obtain the realtime performance while multiplying a matrix by a vector. One can note that solving a system of equations is a part of solving the sample problem. The task that is carried out in the present subsection is taken up just out of curiosity (or academic interest) since, unlike the task that is concerned with just multiplying a matrix by a vector, present author cannot think of any use in achieving the realtime performance just for the portion of the MATLAB code that just solves the system of simultaneous equations.

One can see that the time needed to multiply a matrix by a vector depends on the size of the matrix, not on the actual values of the elements of the matrix. Similarly, at least when not using an iterative solver, the time needed to solve a system of equations is mainly dependent on the size of the system of equations only. Hence, in the present work, instead of solving the actual system of equations obtained through BEM, a dummy system of equations is generated and solved. Using dummy system of equations is useful here because, while the sample problem always generates a system of equations that has 288 simultaneous equations, different problem sizes can easily be tried out if dummy system of equations are utilized. Results show that when the size of the system of equations is equal to 500, the system of equations can be solved in realtime, either by making use of the GPU or otherwise; here, solving the system of equations on the GPU, and solving the system of equations on the CPU, both take almost the same amount of time. But when the size of the system of equations is equal to around 1000, the system of equations can be solved in realtime only if the GPU is made use of. And when the size of the system of equations is equal to 1500, the system of equations cannot be solved in realtime whether a GPU is used or not.                 

\subsection{Using the GPU to Multiply a Matrix by a Vector}

Motivation for the present attempt to obtain the realtime performance while multiplying a matrix by a vector is explained in the previous subsection above. The same arguments used in the last subsection to make use of dummy problems are applicable to the present subsection too.

From the results, when the size of the vector is 16000, the simulation on the CPU takes 0.418 s while the simulation that uses the GPU takes just 0.030 s (i.e., 14 times faster). The thirty computations per second desired by realtime graphics amounts to an allowable time of up to 0.033 s per computation, and one can note that this targeted speed for this simulation cannot be achieved if the GPU is not made use of here.
    
\subsection{Running the Sample Problem on a Computer Cluster}

This subsection is a very important part of the present paper. Motivation for trying to obtain the realtime performance while solving the whole of the sample problem on a cluster has already been explained in the first section (i.e., `Introduction') of the present paper; also, the last section (i.e., `Concluding Remarks') mentions some points related to the present subsection.

In this subsection, whole of the sample problem (not a part) is run on a cluster. The sample problem is solved using 1, 4, 16, 64, and 256 processors, and the time needed to solve the sample problem is noted down for each of the cases. One can note that the code first calculates the `characteristic matrix' and the `right hand side', and then solves the system of equations; the `time' needed to solve the problem (as noted down in the tables included in this subsection) always includes both the time needed to calculate the `characteristic matrix' and the `right hand side' and the time needed to solve the system of equations. Part of the code that calculates the `characteristic matrix' and the `right hand side' is separated from the part of the code that solves the system of equations by using the BLACS routine `blacs\_barrier'; hence, in the code, the task of solving the system of equations begins only just after the whole of both the `characteristic matrix' and the `right hand side' are assembled (i.e., solution of the system of equations can begin only after each of the processes in the process grid complete their part of the work in calculating the `characteristic matrix' and the `right hand side'). Parallelization of the part of the code that calculates the `characteristic matrix' and the `right hand side' uses the `Block Distribution' whereas the part of the code that solves the system of equations uses the `Block-Cyclic Distribution'. When 4 processors are used, a 2 by 2 process grid is used; when 16 processors are used, a 4 by 4 process grid is used; when 64 processors are used, an 8 by 8 process grid is used; and when 256 processors are used, a 16 by 16 process grid is used. Parallel version of the Fortran code that solves the sample problem on the cluster uses ScaLAPACK to solve the system of equations while the sequential version of the Fortran code that solves the sample problem on a single core of the cluster uses LAPACK while solving the system of equations. The parallel version of the Fortran code uses BLACS and MPI also.

One can see that the present simulation uses a `characteristic matrix' of 288 by 288 size. One can note that when the parallelized Fortran code is run on a single processor, the `characteristic matrix' is Block-Cyclically distributed by using a block size of 288 by 288. But here, when the parallelized Fortran code is run on 4 processors, block sizes of 144, 128, 64, 32, and 1 are tried out. Similarly, when the code is run on 16 processors, block sizes of 64, 32, and 1 are tried out; when the code is run on 64 processors, block sizes of 32 and 1 are tried out; and when the code is run on 256 processors, a block size of 16 is used. Idea behind choosing these block sizes is that, from literature, codes execute slower if the block sizes are too small (like 1), and again, if the block sizes are too large, some of the processors may not get any data to process and hence the very use of higher number of processors to achieve better parallelism may lose its purpose; also, some references recommend using a block size of 32, 64, or even 128, for good performance. Hence the block sizes for different cases that use different number of processors are chosen such that all the processors get some data to process, and different block sizes are tried out for the same cases to find out how block sizes affect the speed.

Now, the time taken for the code to execute on different number of processors, for different block sizes, is listed in Table 1; results are presented for four trials, and the averages of the four trials are also listed. Now, Table 2 gives the average time taken by the code to execute itself on different total number of processors, taking into account all the block sizes considered for the case, and taking into account all the trials of a particular simulation also.

\begin{table}[htbp]
\centering
\caption{Solution Time in Seconds}
\begin{tabular}{lccccc}
\hline
& First Run & Second Run & Third Run  & Fourth Run & Average \\   
    Serial (with `-mkl=sequential') & 0.170 & 0.246 & 0.165 & 0.134 & 0.179 \\ 
    Threaded (with `-mkl') & 0.364 & 0.335 & 0.330 & 0.246 & 0.319 \\
    Parallel (1 Process, Block Size=288) & 0.340 & 0.230 & 0.149 & 0.182 & 0.225 \\
    Parallel (4 Processes, Block Size=144) & 0.226 & 0.069 & 0.136 & 0.053 & 0.121 \\
    Parallel (4 Processes, Block Size=128) & 0.219 & 0.116 & 0.099 & 0.106 & 0.135 \\
    Parallel (4 Processes, Block Size=64) & 0.406 & 0.497 & 0.212 & 0.460 & 0.394 \\
    Parallel (4 Processes, Block Size=32) & 0.221 & 0.187 & 0.255 & 0.239 & 0.225 \\
    Parallel (4 Processes, Block Size=1) & 0.532 & 0.441 & 0.344 & 0.527 & 0.461 \\
    Parallel (16 Processes, Block Size=64) & 0.055 & 0.067 & 0.063 & 0.070 & 0.064 \\
    Parallel (16 Processes, Block Size=32) & 0.065 & 0.057 & 0.050 & 0.042 & 0.054 \\
    Parallel (16 Processes, Block Size=1) & 0.053 & 0.049 & 0.058 & 0.064 & 0.056 \\
    Parallel (64 Processes, Block Size=32) & 1.008 & 1.229 & 0.750 & 0.078 & 0.766 \\
    Parallel (64 Processes, Block Size=1) & 1.740 & 2.041 & 2.400 & 1.171 & 1.838 \\
    Parallel (256 Processes, Block Size=16) & 2.067 & 1.167 & 1.321 & 1.460 & 1.504 \\
\hline
\end{tabular}
\end{table}

\begin{table}[htbp]
\centering
\caption{Average Solution Time in Seconds, Considering all the Runs and all the Block Sizes that are Considered}
\begin{tabular}{lc}
\hline
Serial (with `-mkl=sequential') & 0.179 \\
    Threaded (with `-mkl') & 0.319 \\
    Parallel (1 Process) & 0.225 \\
    Parallel (4 Processes) & 0.267 \\
    Parallel (16 Processes) & 0.058 \\
    Parallel (64 Processes) & 1.302 \\
    Parallel (256 Processes) & 1.504 \\
\hline
\end{tabular}
\end{table}

From the results presented in Table 1 and Table 2, for the sample problem (i.e., the problem considered here), block sizes do not play any significant and meaningful role in the overall sense. Also, on an average, the speediest performance is obtained when 16 processors together with a block size of 32 are used; for this case, the simulation took 0.054 s to complete (this corresponds to about 19 computations per second). One can also observe that the fastest performance recorded in the tables corresponds to the `Fourth run' corresponding to the case when 16 processors together with a block size of 32 are used; for this instance, the simulation took 0.042 s to complete (and this corresponds to about 24 computations per second). And one can observe that if the average is taken for all the runs and all the block sizes together, on an average, speediest performance is obtained when 16 processors are used (solution time = 0.058 s; this corresponds to about 17 computations per second).

One might need to keep a few points in mind while studying the results presented in this subsection. The cluster used here is made up of heterogeneous nodes. The nodes with 32 cores have 2.4 GHz processors whereas the nodes with 64 cores have 2.2 GHz processors. Also, internode communications could be slower when compared to intranode communications. In the present subsection, simulations using 1, 4, and 16 processors are usually run on a node having 32 cores, and the simulation that uses 64 processors is run on a node having 64 cores; the simulation requiring 256 processors is run using four nodes with 64 cores each. Of course, for each and every simulation mentioned in this subsection, each of the processors runs one and only one process.

One can see that one is not likely to achieve faster performance by going for higher number of processors in this cluster.

One can note that because the whole of the sample problem is made to run on a cluster here, results presented here are useful when one wishes to learn about the performance of BEM when the `characteristic matrix' changes during simulations (e.g., during the simulation of cutting, during the simulations that use nonlinear BEM). One can also note that the BEM used here is the `standard' BEM, not any specialized version of BEM (e.g., the Fast Multipole Boundary Element Method which uses the fast multipole method to accelerate the solution of the system of equations).   

\subsection{Possibility of Simulating Nonlinear Behavior in Realtime using BEM}

Many a times, realistic description of nonlinear behavior of biological organs (like liver) requires the use of hyperelastic material models (e.g., Mooney-Rivlin model, Neo-Hookean model). Solving one hyperelastic problem is equivalent to solving many linear problems. Although the total number of iterations (i.e., linear solutions or Newton iterations) required to solve a nonlinear problem within the specified tolerance (for the error) cannot be known beforehand, one can get an idea of the total number of iterations needed to solve a nonlinear problem by referring to the literature that deals with the solution of similar type of problems. For example, [43] includes the task of solving a hyperelastic problem, and by going through [43], the total number of Newton iterations needed to solve the problem is always between 62 to 70. Also, by making use of software like ANSYS which have in-built hyperelastic material models, one can solve similar hyperelastic problems to get an idea of the total number of iterations needed to solve such problems. Thus by solving dummy hyperelastic problems on ANSYS, and also by referring to the literature dealing with the solution of hyperelastic problems, one can see that solving a hyperelastic problem takes about 5, 10, 20, 50 or even 80 Newton iterations, while solving different nonlinear problems; it is also observed by the present author that more than 100 linear solutions are rarely needed to solve a hyperelastic problem, although this conclusion is reached just by observing a limited number of examples. Hence it is reasonable to assume that a hyperelastic problem can be solved in realtime if the corresponding linear problem can be solved in realtime 100 times.

But looking at the results presented in the previous subsections of this section, it may be difficult to obtain the realtime performance with a hyperelastic material model.
   
\section{Concluding Remarks}

In this work, a very brief description of some BEM codes developed by the present author is given first. The codes deserve mention in the paper since the codes are necessary to carry out the present work. The codes comprise a MATLAB code that can solve any linear elastostatic problem without accounting for body forces, and a Fortran translation of the MATLAB code. The Fortran translation comes in two varieties: first one can run on a single core of a computer cluster (i.e., sequential version), and the second variety can utilize multiple processors available in computer clusters (i.e., parallelized version). 

Next, a try is given to parallelize the entire MATLAB code with the intention of running it on a desktop computer equipped with a GPU. Here the goal is to compute the `characteristic matrix' and the `right hand side', and also to solve the system of equations, all in realtime. Author has not been successful in achieving this goal using MATLAB and the Parallel Computing Toolbox.

Next, thought was given to obtain the realtime performance, again by utilizing just an ordinary desktop computer and a GPU, by resorting to precomputations. In this case, one has to precompute the `characteristic matrix' and its inverse offline. Again, MATLAB and Parallel Computing Toolbox are used for this purpose. Results show that it is possible to get realtime performance (realtime graphics, not realtime haptics) when the size of the `characteristic matrix' is up to about 16000 by 16000; one can also see that one can get a speed up of about 14 times when the `characteristic matrix' is of about 16000 by 16000 size. This means that one can perform realtime simulations using boundary elements if the geometry is described by about 5300 constant boundary elements. Of course, the usual limitations with approaches that use precomputations apply in this case too (like one may not be able to compute the `characteristic matrix' beforehand if there is change in the geometry during simulations, e.g., during simulation of cutting). But one can note that it is not possible to perform the same simulation (with the same precomputions of course) in realtime using just an ordinary desktop if a GPU is also not made use of.

Next, a computer cluster is used to carry out BEM simulations. The fully parallelized version of the code (i.e., the parallel Fortran code) is used for this purpose. In this case, the goal is to compute the `characteristic matrix' and the `right hand side', and also to solve the system of equations, all in realtime. Simulations are carried out on 1, 4, 16, 64, and 256 processors. Also, simulations are carried out for different block sizes while Block-Cyclically distributing matrices among processors. Results indicate that one can barely obtain a good realtime performance as far as realtime graphics is concerned; the fastest simulation could complete 24 computations per second (as against the required 30 computations per second for high quality realtime graphics). One has to note that the present study has used only standard methods of parallelizing a program; no custom methods and communication protocols have been used. Although using a supercomputer (e.g., IBM Blue Gene) instead of the computer cluster utilized in the present work may offer slightly different speed/performance, one may not expect the speed to improve too much because the well known list of top supercomputers of the world includes several computer clusters at present (which implies that the performance offered by clusters is comparable to the performance offered by supercomputers). Using an 8 by 8 integration scheme instead of the 16 by 16 integration scheme employed in the present study can speed up the computations, which in turn can enable one to achieve realtime graphics of very high quality and/or enable one to solve a problem of larger size in realtime; author has observed that using 8 by 8 integration instead of the 16 by 16 integration does not result in unacceptable degradation of accuracy. Also, present study carries out all the computations involving real numbers in double precision, and carrying out at least a few computations using just the single precision can help to speed up the computations.  

One can see that the results presented in the present paper are relevant not only for the realtime linear elastostatic simulation of biological organs but any simulation that attempts to simulate linear elastostatic response in realtime.

From the results presented in the paper, one can see that it would be difficult to obtain the realtime performance (even realtime graphics) if nonlinear material behavior is to be incorporated into simulations, if custom methods and communication protocols are not used while parallelizing simulations. Results also imply that it may be difficult to obtain realtime haptic feedback with nonlinear material models, even if custom methods and communication protocols are used while parallelizing simulations. However, results also indicate that BEM could be a very useful numerical technique for the realistic and realtime simulation of biological organs (including the highly nonlinear soft biological organs) if one is happy with nearly realtime performance, i.e., one computation taking just a few seconds.

Of course, simulations may not need to be strictly realtime sometimes (i.e., a `hard' realtime system may not be needed, but `firm' or `soft' realtime systems may also be suitable). For example, [18] mentions that a neurosurgeon takes about 0.25 seconds to perform each incremental stage of surgical cutting in practice; hence it may be all right for the simulations to take 0.25 seconds to complete in this case. Also, [44] mentions that the allowable lag time is 0.1 seconds for inexperienced users and it is 1 second for experienced users; hence the simulations can take 0.1 or 1 second in this case. References [45-47] mention that the just noticeable difference of the human sensory system for force perception at human fingertips is about 10

Present work also serves to make some record of the performance (like speed) that can be obtained by present day typical computing hardware (like a desktop computer, a graphics processing unit (GPU), a computer cluster) together with contemporary software (like MATLAB, MATLAB Parallel Computing Toolbox, Fortran, MPI, BLACS, ScaLAPACK).

As to the limitations of the present work, only `standard' methods of parallelization are tried out; only standard and readily available tools like BLACS, ScaLAPACK, MPI are used. But this in fact is done purposefully, with the intention of knowing whether one can achieve the targeted realtime performance without using custom methods and communication protocols; the task of parallelizing should not itself be a research problem. The second limitation could be that the paper uses just a simple geometry (i.e., a cantilever beam) to demonstrate the speed of BEM, and does not mention anything about the speed of BEM when simulations are carried out on complicated geometries like biological organs. But one can note that the speed of a simulation here depends solely on the total number of elements and boundary conditions, and the speed does not have any relevance to the actual geometry; hence there is no need to carry out simulations on different geometries just to measure how fast (or slow) the simulations are. The third limitation of the present study could be that only one problem is taken up here for studying the performance of the parallel Fortran code on the computer cluster, and different problem sizes have not been tried out on the cluster to understand how the speed varies with problem size. But since there is no reason why a larger sized problem would execute faster when compared to a smaller sized problem, and since the problem that has been taken up and solved can be solved only for a maximum of 24 times a second, there really is not a need to go for a larger problem size as far as the realtime performance is concerned although going for a larger problem size may offer better performance in the sense that there may be only a small increase in the solution time when the problem size increases by a significant amount. In fact, author has executed a larger sized problem on the same computer cluster using different number of processors and for different block sizes but the problem could never be solved more than 24 times a second for any of the trials considered.  

Future work could be to develop an exhaustive BEM library that could be used to solve highly nonlinear problems on parallel computers. The library could then be utilized to perform either realtime or nearly realtime simulation of biological organs. Finally the simulations could be integrated into a surgical simulator.

\section*{Acknowledgments}

Author is grateful to the Robotics Lab, Department of Mechanical Engineering \& Centre for Product Design and Manufacturing, Indian Institute of Science, Bangalore, INDIA, for providing access to a Graphics Processing Unit (GPU). \\ \\
Author acknowledges Supercomputer Education and Research Centre (SERC), Indian Institute of Science, Bangalore, INDIA, for providing access to computer clusters. \\ \\
Author is grateful to the Centre for Product Design and Manufacturing, Indian Institute of Science, Bangalore, INDIA, for providing an opportunity to carry out the present work. \\ \\
Author is indebted to the Indian Institute of Science, Bangalore, INDIA, for providing the necessary infrastructure to carry out the present work.

\section*{References}
 
\verb|[1]| U. Meier, O. Lopez, C. Monserrat, M. C. Juan, M. Alcaniz, 2005, Real-time deformable models for surgery simulation: a survey, Computer Methods and Programs in Biomedicine, 77(2005), pp. 183-197, doi:10.1016/j.cmpb.2004.11.002 \\ \\
\verb|[2]| Herve Delingette, 1998, Toward Realistic Soft-Tissue Modeling in Medical Simulation, Proceedings of the IEEE, Vol. 86, No. 3, March 1998, pp. 512-523. \\ \\
\verb|[3]| Herve Delingette, Nicholas Ayache, Soft Tissue Modeling for Surgery Simulation, Online resource from \url{http://www.inria.fr/centre/sophia/} (accessed October 25, 2013) \\ \\
\verb|[4]| Yongmin Zhong, Bijan Shirinzadeh, Julian Smith, Chengfan Gu, 2009, An electromechanical based deformable model for soft tissue simulation, Artificial Intelligence in Medicine, 47(2009), pp. 275-288, doi:10.1016/j.artmed.2009.08.003 \\ \\
\verb|[5]| Sarthak Misra, K. T. Ramesh, Allison M. Okamura, 2008, Modeling of Tool-Tissue Interactions for Computer-Based Surgical Simulation: A Literature Review, Presence, October 2008, Vol. 17, No. 5, pp. 463-491. \\ \\
\verb|[6]| Jung Kim, Mandayam A. Srinivasan, 2005, Characterization of viscoelastic soft tissue properties from in vivo animal experiments and inverse FE parameter estimation, Medical Image Computing and Computer-Assisted Intervention - MICCAI 2005, Lecture Notes in Computer Science, Volume 3750, pp. 599-606. \\ \\
\verb|[7]| Jung Kim, Bummo Ahn, Suvranu De, Mandayam A. Srinivasan, 2008, An efficient soft tissue characterization algorithm from in vivo indentation experiments for medical simulation, The International Journal of Medical Robotics and Computer Assisted Surgery, Volume 4, Issue 3, pp. 277-285, September 2008, DOI: 10.1002/rcs.209 \\ \\
\verb|[8]| Basdogan C., 2001, Real-time simulation of dynamically deformable finite element models using modal analysis and spectral Lanczos decomposition methods, Stud Health Technol Inform. 2001;81:46-52. \\ \\ 
\verb|[9]| Ahn B., Kim J., 2010, Measurement and characterization of soft tissue behavior with surface deformation and force response under large deformations, Medical Image Analysis, 14, pp. 138-148. \\ \\  
\verb|[10]| Firat Dogan, M. Serdar Celebi, 2011, Real-time deformation simulation of non-linear viscoelastic soft tissues, Simulation, Volume 87, Issue 3, March 2011, pp. 179-187, doi:10.1177/0037549710364532 \\ \\ 
\verb|[11]| \url{http://imechanica.org/node/3239} (accessed October 24, 2013) \\ \\ 
\verb|[12]| \url{http://www.stanford.edu/group/sailsbury_robotx/cgi-bin/salisbury_lab/?page_id=343} (accessed October 24, 2013) \\ \\ 
\verb|[13]| \url{http://www.ansys.com/staticassets/ANSYS/staticassets/resourcelibrary/article/AA-V5-I2-Cut-to-the-Bone.pdf} (accessed October 24, 2013) \\ \\ 
\verb|[14]| Kirana Kumara P, Ashitava Ghosal, 2012, Real-time Computer Simulation of Three Dimensional Elastostatics using the Finite Point Method, Applied Mechanics and Materials, 110-16, pp. 2740-2745. \\ \\ 
\verb|[15]| Suvranu De, Jung Kim, Yi-Je Lim, Mandayam A. Srinivasan, The point collocation-based method of finite spheres (PCMFS) for real time surgery simulation, Computers and Structures, 83(2005), pp. 1515-1525. \\ \\ 
\verb|[16]| Ullrich Meier, Carlos Monserrat, Nils-Christian Parr, Francisco Javier Garcia, Jose Antonio Gil, 2001, Real-Time Simulation of Minimally-Invasive Surgery with Cutting Based on Boundary Element Methods, Medical Image Computing and Computer-Assisted Intervention - MICCAI 2001, Lecture Notes in Computer Science, Volume 2208, pp. 1263-1264. \\ \\ 
\verb|[17]| C. Monserrat, U. Meier, M. Alcaniz, F. Chinesta, M.C. Juan, A new approach for the real-time simulation of tissue deformations in surgery simulation, Computer Methods and Programs in Biomedicine, 64(2001), pp. 77-85. \\ \\ 
\verb|[18]| P. Wang, A.A. Becker, I.A. Jones, A.T. Glover, S.D. Benford, C.M. Greenhalgh, M. Vloeberghs, Virtual reality simulation of surgery with haptic feedback based on the boundary element method, Computers and Structures, 85(2007), pp. 331-339. \\ \\ 
\verb|[19]| Foster, Timothy,Mark, 2013, Real-time stress analysis of three-dimensional boundary element problems with continuously updating geometry, Doctoral thesis, Durham University. \\ \\ 
\verb|[20]| Robert G Aykroyd, Brian A Cattle, Robert M West, A boundary element approach for real-time monitoring and control from electrical resistance tomographic data, 4th World Congress on Industrial Process Tomography, Aizu, Japan. \\ \\ 
\verb|[21]| P Wang, A A Becker, I A Jones, A T Glover, S D Benford, M Vloeberghs, 2009, Real-time surgical simulation for deformable soft-tissue objects with a tumour using Boundary Element techniques, Journal of Physics: Conference Series, Volume 181, Number 1, doi:10.1088/1742-6596/181/1/012016 \\ \\ 
\verb|[22]| Bo Zhu, Lixu Gu, 2012, A hybrid deformable model for real-time surgical simulation, Computerized Medical Imaging and Graphics, 36(2012), pp. 356-365. \\ \\ 
\verb|[23]| Martin Bayliss, 2003, The Numerical Modelling of Elastomers, PhD Thesis, School of Engineering, Cranfield University. \\ \\   
\verb|[24]| Alex Rhomberg, 2001, Real-time finite elements: A parallel computer application, Doctoral Thesis, Swiss Federal Institue of Technology, Zurich, Publisher: Shaker, http://dx.doi.org/10.3929/ethz-a-004089534 \\ \\ 
\verb|[25]| M. Kreienmeyer, E. Stein, 1995, Parallel implementation of the boundary element method for linear elastic problems on a MIMD parallel computer, Computational Mechanics, 15(1995), pp. 342-349. \\ \\ 
\verb|[26]| Kirana Kumara P, 2014, Codes for solving three dimensional linear elastostatic problems using constant boundary elements while ignoring body forces, Available from: \url{http://eprints.iisc.ernet.in/48088/} (accessed January 10, 2014) \\ \\ 
\verb|[27]| W. T. Ang, 2007, A Beginner's Course in Boundary Element Methods, Universal Publishers, Boca Raton, USA. \\ \\ 
\verb|[28]| Gernot Beer, Ian Moffat Smith, Christian Duenser, 2008, The Boundary Element Method with Programming: For Engineers and Scientists, Springer \\ \\ 
\verb|[29]| C.A. Brebbia, J. Dominguez, Boundary Elements - An Introductory Course, Second Edition, WIT Press \\ \\ 
\verb|[30]| Brebbia CA, 1978, The boundary element method for engineers, London/New York, Pentech Press/Halstead Press \\ \\ 
\verb|[31]| Youssef F. Rashed, Available from: \url{http://www.boundaryelements.com/} (accessed October 22, 2013) \\ \\ 
\verb|[32]| M. Guiggiani, A. Gigante, 1990, A General Algorithm for Multidimensional Cauchy Principal Value Integrals in the Boundary Element Method, Journal of Applied Mechanics, Transactions of the ASME, Vol. 57, December 1990, pp. 906-915. \\ \\ 
\verb|[33]| M. Guiggiani, Formulation and numerical treatment of boundary integral equations with hypersingular kernels, Contained in: ``Singular Integrals in B. E. Methods, V. Sladek and J. Sladek, eds., 1998'' \\ \\ 
\verb|[34]| \url{http://www.efunda.com} (accessed October 22, 2013) \\ \\ 
\verb|[35]| Errata, \url{http://www.ntu.edu.sg/home/mwtang/bem2011.html#errata} (accessed October 22, 2013) \\ \\ 
\verb|[36]| Alastair McKinstry, Alin Elena, Ruairi Nestor, An Introduction to Fortran, Available from: \url{www.ichec.ie/support/tutorials/fortran.pdf} (accessed October 22, 2013) \\ \\ 
\verb|[37]| Blaise Barney, Message Passing Interface (MPI), Available from: \url{https://computing.llnl.gov/tutorials/mpi/} (accessed October 22, 2013) \\ \\ 
\verb|[38]| Intel Math Kernel Library Reference Manual, Available from: \url{http://software.intel.com/sites/products/documentation/hpc/mkl/mklman/index.htm} (accessed October 22, 2013) \\ \\ 
\verb|[39]| ScaLAPACK Example Programs, Available from: \url{http://www.netlib.org/scalapack/examples/} (accessed October 22, 2013) \\ \\ 
\verb|[40]| ScaLAPACK - Scalable Linear Algebra PACKage, Available from: \url{http://www.netlib.org/scalapack/} (accessed October 22, 2013) \\ \\ 
\verb|[41]| Parallel ESSL Guide and Reference, Available from: \url{http://publib.boulder.ibm.com/infocenter/clresctr/vxrx/} (accessed November 29, 2013) \\ \\
\verb|[42]| Simulation Acceleration using System Objects, MATLAB Coder and Parallel Computing Toolbox, Documentation Center, MathWorks, Available from: \url{http://www.mathworks.in/help/comm/examples/simulation-acceleration-using-system-objects-matlab-coder-and-parallel-computing-toolbox.html} (accessed October 30, 2013) \\ \\ 
\verb|[43]| Mark Adams, James W. Demmel, 1999, Parallel Multigrid Solver for 3D Unstructured Finite Element Problems, ACM 1-58113-091-8/99/0011, Available from: \url{www.cs.illinois.edu/~snir/PPP/mesh/multigrid3.pdf} (accessed October 30, 2013) \\ \\  
\verb|[44]| Valerie E. Taylor, Milana Huang, Thomas Canfield, Rick Stevens, Daniel Reed, Stephen Lamm, Performance Modeling of Interactive, Immersive Virtual Environments for Finite Element Simulations, International Journal of High Performance Computing Applications, June 1996, vol. 10, no. 2-3, pp. 145-156, doi: 10.1177/109434209601000203 \\ \\ 
\verb|[45]| Sonya Allin, Yoky Matsuoka, Roberta Klatzky, 2002, Measuring Just Noticeable Differences For Haptic Force Feedback: Implications for Rehabilitation, Proceedings of the 10th Symposium on Haptic Interfaces for Virtual Environment and Teleoperator Systems (HAPTICS'02), IEEE Computer Society \\ \\ 
\verb|[46]| H. Pongrac, B. Farber, P. Hinterseer, J. Kammerl, E. Steinbach, 2006, Limitations of human 3D-force discrimination, Human-Centered Robotics Systems \\ \\ 
\verb|[47]| Hong Z. Tan, Brian Eberman, Mandayam A. Srinivasan, Belinda Cheng, 1994, Human Factors for the Design of Force-reflecting Haptic Interfaces, DSC-Vol.55-1, Dynamic Systems and Control, Editor: Clark J. Radcliffe, Book No. G0909A-1994, The American Society of Mechanical Engineers

\section*{Appendix: Explanation of Some Terminologies}

\textbf{BLACS:} The BLACS (Basic Linear Algebra Communication Subprograms) is a linear algebra oriented message passing interface that is implemented efficiently and uniformly across a large range of distributed memory platforms. \\ \\
\textbf{blacs\_barrier:} The `blacs\_barrier' is a routine that holds up execution of all processes until all the processes have called the routine. \\ \\
\textbf{Block Distribution, and Block-Cyclic Distribution:} The two types of data distribution are explained here with reference to a one dimensional array; the explanations presented here for the vector in one dimension can be applied independently over two dimensions (for the rows and columns of a matrix) and that results in a matrix being distributed among processes as per the concerned distribution. Block Distribution assigns blocks of size $r$ of the global vector of $M$ data objects over $P$ processes. For block distribution, the mapping $m \to (p, i)$ is defined as: $m \to (floor(m/L), m \bmod L)$, where $L = ceiling (M/P)$, $m$ is the global index of a data object $(0 \le m < M)$, $p (0 \le p < P)$ specifies the process to which the data object is mapped, and $i$ specifies its location in the local array. In the Block-Cyclic Distribution, blocks of $r$ consecutive data objects are distributed cyclically over the $P$ processes. This can be described by a mapping of the global index $m$ to an index triplet $(p,b,i)$, where $b$ is the block number in process $p$, and other symbols have the same meaning as earlier. Hence, for block-cyclic distribution, the mapping $m \to (p, b, i)$ is defined as: $m \to (floor((m \bmod T)/r), floor(m/T), m \bmod r)$, where $T = rP$. \\ \\
\textbf{Block size:} When a matrix or a vector is to be Block-Cyclically distributed among the available processes, one needs to divide the matrix or the vector into pieces. The number of rows in a piece is called the `row block size' while the number of columns in the piece is called the `column block size'. \\ \\
\textbf{CUDA:} CUDA (Compute Unified Device Architecture) is a parallel computing platform and programming model used by NVIDIA (NVIDIA is an industry leader in the visual computing field). CUDA is very useful whenever GPUs are used for performing scientific computations. \\ \\
\textbf{maxNumCompThreads:} The MATLAB command `maxNumCompThreads' specifies the number of threads that MATLAB can use to run MATLAB. The maximum value that `maxNumCompThreads' can take is equal to the number of computational cores on any machine. \\ \\  
\textbf{MPI:} Message Passing Interface (MPI) is a standardized and portable message-passing system designed to function on a wide variety of parallel computers. \\ \\
\textbf{Parallel Computing Toolbox:} The MATLAB Parallel Computing Toolbox enables some of the MATLAB codes and scripts to be executed on a GPU. \\ \\
\textbf{parallel.gpu.gpuArray.zeros:} The `parallel.gpu.gpuArray.zeros' is a command available through the MATLAB Parallel Computing Toolbox. The command is used to initialize an array in the GPU memory with zeros. \\ \\
\textbf{parfor, matlabpool, MATLAB client, MATLAB lab, MATLAB worker, matlabpool close:} These terminologies are related to the MATLAB Parallel Computing Toolbox, specifically when one wants to utilize multiple cores available in computers. The `parfor' loop is a replacement for the standard MATLAB `for' loop when one wants to run the statements inside the `for' loop on a number of cores in parallel. The command `matlabpool' enables the parallel language features in the MATLAB language by starting a parallel job that connects the MATLAB client with a number of MATLAB labs; the process which starts the parallel environment is called `MATLAB client' while the other processes in the parallel environment are called `MATLAB labs'. MATLAB labs are also known as `MATLAB workers'. The command `matlabpool close' disables the parallel language features in the MATLAB and closes the parallel environment. \\ \\
\textbf{Process grid:} Although the processes of a parallel computer are many a times imagined as a one-dimensional array of processes, sometimes it is more convenient to map this one-dimensional array of processes into a two-dimensional rectangular grid; this rectangular grid is called the process grid. The number of rows in a process grid when multiplied by the number of columns in the process grid gives the total number of processes. \\ \\
\textbf{ScaLAPACK:} ScaLAPACK (Scalable Linear Algebra PACKage) is a library of high-performance linear algebra routines for parallel distributed memory computers. 

\end{document}